\documentstyle[11pt]{article}
\begin{document}
\pagenumbering{arabic}

\def\Bbb{\bf}

\begin{flushright}
\renewcommand{\textfraction}{0}
April 28th 1995\\
hep-th/9504149\\
PEG-05-95\\
\end{flushright}

\begin{center}
{\LARGE {\bf Symmetry in the Topological Phase of String Theory} }
\end{center}
\begin{center}
{\Large Phil Gibbs} \\
e-mail to phil@galilee.eurocontrol.fr
\end{center}

\begin{abstract}
I present arguments to the affect that the topological phase
of string theory must be event-symmetric. This motivates a
search for a universal string group for discrete strings in
event-symmetric space-time which unifies space-time symmetry
with internal gauge symmetry. This is partially successful but
the results are incomplete and I speculate on the use of quantum
groups to define a well behaved theory which would resolve the
discrete/continuum dual nature of stringy space-time.
\end{abstract}


{ \small This document is Copyright \copyright 1995 by the author Philip E.
Gibbs (phil@galilee.eurocontrol.fr). All rights are reserved. Permission to
use, copy and distribute this document by any means and for any
purpose except profit is hereby granted, provided that both the
above copyright notice and this permission notice appear in all copies.
Reproduction in part or in whole by any means, including, but not
limited to, printing, copying existing prints, publishing by electronic or
other means, implies full agreement to the above non-profit-use clause,
unless upon explicit prior written permission of the author. }

\pagebreak

\section*{Introduction}

Despite the lack of experimental data above the Electro-Weak energy scale, the
search for unified theories of particle physics beyond the standard model has
yielded many mathematical results based purely on constraints of high symmetry,
renormalisability and cancellation of anomalies. In particular, space-time
supersymmetry \cite{WeZu74} has been found to improve perturbative
behaviour and to bring the gravitational force into particle physics. One
ambitious but popular line of quantum gravity research is superstring theory
\cite{GrSc81a,GrSc81b,GrSc82}. String models were originally constructed
in perturbative form and were found to be finite at each order \cite{Man92}
but incomplete in the sense that the perturbative series were not Borel
summable \cite{GrPe88}.

Despite this there has been a huge amount of interest in a number
of super-string theories and in the Heterotic String in particular
\cite{GrHaMaRo85}. The fact that this theory has an almost unique
formulation with the interesting gauge group $E_8 \otimes E_8$ persaudes
many that it is the sought after unified field theory despite the fact
that it is only finite in ten dimensional space-time. In 1926 Klein
\cite{Kle26} proposed that a 5 dimensional theory due to Kaluza \cite{Kal21}
could make physical sense if one of the dimensions was compactified.
Kaluza-Klein theory has been applied to the heterotic string theory for which
it has been shown \cite{CaHoStWi85} that six of the ten dimensions could be
spontaneously compactified on a Calabi-Yau manifold or orbifold. This leaves
an $E_6$ gauge group with suitable chirality modes just big enough to
accommodate low energy particle physics. The difficulty which remains
is that there are many topologically different ways the compactification
could happen and there is no known way of picking the right one. To solve
this problem it is thought necessary to find some non-perturbative analysis
of the string theories. As a first step it might be necessary to construct
a second quantised covariant String Field Theory \cite{KaKi74}.

There has been some preliminary success in formulating both open \cite{Wit86}
and closed \cite{Zwi92} bosonic String Field Theories.
There have also been some important steps taken towards background independent
formulations of these theories \cite{Wit92b,SeZw93}.
However, they still fail to provide an explicit unification of
space time diffeomorphism symmetry with the internal gauge symmetry. This is a
significant failure because string theories are supposed to unify gravity
with the other gauge forces and there is evidence that string theory does
include such a unification \cite{HoHoSt91}. Furthermore these formulations
have not yet been extended to superstring theories and this appears to be a
very difficult problem \cite{Ced94}.

Conventional wisdom among the pioneers of string theories was that there
is a unique string theory which is self consistent and which explains all
physics. This view was gradually tempered by the discovery of a variety of
different string theories but recently the belief in uniqueness has been
reaffirmed with the discovery that there are hierarchies in which some
string theories can be seen as contained within others
\cite{BeVa93,BeFrWe93,KuSaTo94b}. This inspires a search for a universal
string theory \cite{Fig93,BaNoPe94,KuSaTo94a}.

A successful theory of Quantum Gravity should describe physics at the
Planck scale \cite{Pla99}. It is likely that there is a phase
transition in string theories at their Hagedorn temperature near $kT =$ Planck
Energy \cite{Hag68}. It has been speculated that above
this temperature there are fewer degrees of freedom and a restoration of a much
larger symmetry \cite{AtWi88,GrMe87,GrMe88,Gro88}. This phase is sometimes
known as the topological phase because it is believed that a Topological
Quantum Field Theory may describe it. A fundamental formulation of
string theory would be a model in which the large symmetry is explicit.
It would reduce to the known formulations after spontaneous symmetry
breaking below the Hagedorn Temperature.

One interpretation of the present state of string theories is that it
lacks a geometric foundation and that this is an obstacle to finding
its most natural formulation. It is possible that our concept of
space-time will have to be generalised to some form of ``stringy space''
in which its full symmetry is manifest. Such space-time must be
dynamical and capable of undergoing topological or even dimensional changes
\cite{AsGrMo93,AnFeKo94,GrMoSt95}. To understand stringy space it is almost
certainly necessary to identify the symmetry which is restored at high
temperature. I suggest that the symmetry should include event-symmetry of
space-time \cite{Gib94a}.


\section*{Event-Symmetric String Field Theory}

The fact that a large number of degrees of freedom are perhaps required to
produce event-symmetry breaking suggests that string theories might
provide suitable models. We know that random matrix models are
pre-theories for string theory at $c \leq 1$. In matrix models the size
$N$ of the matrix is taken to infinity in a scaling limit. $N$ is usually
interpreted to be a number of colours in a gauge theory. However, there is
an alternative interpretation that $N$ is the number of space-time events
and that a one-matrix model can be regarded as a pregeometric model of
space-time. Apparently this interpretation was suggested by Dyson at least
ten years ago as stated in \cite{KaWe85}. In this case a matrix model is
event-symmetric in the sense defined by myself \cite{Gib94a} which suggests
that other similar event-symmetric models may also be pre-theories for strings.

The place to start our search for other models is
with Kaku's version of string groups \cite{Kak88a,Kak88b}.

To define the string groups an open oriented string is first considered as
a topological object independently of any target space. There is an abstract
operator $L_C$ for each string $C$ and a Lie-product is defined for these
operators by specifying the structure constants,
\begin{equation}
                 [ L_A, L_B ] = \sum f^C_{AB} L_C
\end{equation}
Three strings $(A,B,C)$ are said to form a triplet if the end of A matches
the start of B, the end of B matches the start of C and the end of C matches
the start of A. They must match in such a way that there is no part of any
string unmatched. In other words they form three matching lengths
radiating from one central point.

When $(A,B,C)$ is such a triplet then the structure constants are
\begin{equation}
                 f^{C^T}_{AB} = f^{A^T}_{BC} = f^{B^T}_{CA} = 1
\end{equation}
and
\begin{equation}
                 f^{C^T}_{BA} = f^{A^T}_{CB} = f^{B^T}_{AC} = -1
\end{equation}
$C^T$ is the transposed string formed by reversing its orientation.

All other structure constants are zero. It can be checked that this
does define a Lie-algebra because the product is anti-symmetric
and satisfies the Jacobi identity. This Lie-algebra is regarded as generating
the gauge group of the open string field theories.

The algebra can be realised on a continuous $D$ dimensional manifold $M_D$
when the strings are continuous open oriented curve segments in the space.
The group is then called the Universal String Group. Such string groups have
been used in attempts to formulate both open and closed string field theories
\cite{Kak90,Kak91}.

These formulations are, however, not completely satisfying. If string
theory really unifies gravity with the gauge forces then the symmetry
group of gravity, $diff(X)$ on a manifold $X$, should be unified with the
string group, $str(X)$ on target space $X$. This is not achieved in the
continuum string field theories. Furthermore the presence of topology change
and mirror manifolds suggests that it is not possible \cite{Wit93} because
the string groups must be the same on topologically different manifolds.
Ideally the diffeomorphism  groups should be contained in the full string
groups
\begin{equation}
                 diff(X) \subset str(X) \simeq str(X')
\end{equation}
But then the string group must contain $diff(X)$ and $diff(X')$ for
topologically different manifolds $X$ and $X'$.

This seems at first quite unreasonable but in fact it is exactly what happens
in event-symmetric field theories which contain the full event-symmetric
group $S(X)$. These groups are isomorphic for any two manifolds and contain
the diffeomorphism groups
\begin{equation}
                 diff(X) \subset S(X) \simeq S(X')
\end{equation}
The solution will be to find a string group which contains the symmetric group
\begin{equation}
                 diff(X) \subset S(X) \subset str(X)
\end{equation}
It seems that the phenomena of topology change is telling us that string
theory must be event-symmetric. If this is true then it is natural to
speculate that event-symmetry is part of the larger symmetry which is restored
above the Hagedorn temperature. It could be said that space-time itself
evaporates at this temperature.

Before we jump to too many conclusions about the phase structure of string
theory it is worth remembering that matter density is as important as
temperature in critical phenomenology. The Hagedorn phase transition seems
to be analogous to evaporation of a liquid forming a string gas. In the
density-temperature phase diagram for water it is possible to go from
the gas phase to the liquid phase at high density without ever passing
through a phase change. If this is also true for string theory then we
cannot really say that the event symmetry is broken in the low temperature
(liquid) phase. It is still there bit somehow less evident. Until we
have a full non-perturbative formulation of string theory it is difficult
to know what the phase diagram is really like.

Is there an inconsistency between the conventional view that the
high temperature state of string theory is a topological phase and
this new idea that it is event symmetric? Not necessarily. Remember
that in a topological quantum theory the relationship between two different
points in space-time is independent of where they are on the
manifold. This is also true in an event symmetric model. But, TQFT is
not event symmetric, it has a diffeomorphism structure. However, a TQFT
describes a field theory on a fixed manifold. A complete theory of
quantum gravity should be a suitable weighted sum over all admissible
topologies of space-time. It is quite possible that such a sum could
effectively remove the diffeomorphism structure leaving a totally event
symmetric theory. A one-matrix model gives an example of how this works.
This simple observation is the basis for uncovering the true nature of
string theory.

We shall see how it is possible to define string groups on an
event-symmetric target space of discrete points in such a way that the
symmetric group is included as a subgroup. In this way we can achieve
a natural unification of space-time symmetries in the form of the
event-symmetric group and the gauge groups in the form similar to the
Universal String Group.


\section*{Are Strings Discrete?}

An event-symmetric formulation of string theory is bound to be discrete so
it is worth reviewing other peoples ideas about the discreteness of string
theory before commencing.

The notion that string theory has a minimum length is well established as a
result of target space duality which provides a transformation from
distances $R$ to distances $\alpha/R$ where $\sqrt \alpha$ is the size of
compactified dimensions at the Planck scale. A minimum length does not
necessarily imply discrete space-time but it is suggestive.

Thorn has argued that large $N$ matrix models lead to an interpretation of
string theories as composed of pointlike partons \cite{Tho94a,Tho94b}. A
similar view has been pursued by Susskind as a resolution of paradoxes
concerning Lorentz Contraction at high boosts and the black hole information
loss puzzle \cite{Sus93,Sus94}. It is possible to calculate exact string
amplitudes from a lattice theory with a non-zero spacing \cite{KlSu88}.

There are some remarkable features about these discrete string models. Firstly
it is found that when the spacing between discrete partons is reduced below
a certain limit there is a phase transition beyond which results coincide
exactly with continuous models \cite{KlSu88,Dal94,Kos94}. Even
more odd is the apparent generation of an extra dimension of space so
that models in 2+1 dimensions could become theories of 3+1 dimensions
\cite{Rus93,Tho94a,SuGr94}. It seems that these concepts are not inconsistent
with the algebraic construction of Topological Quantum Field Theory
\cite{Cra95}

Perhaps the fact that string theory is finite to each order in perturbation
theory is itself an indication that string theory is discrete. In lattice
theories the renormalisation group is used to send the lattice spacing
to zero but in string theory the coupling is not renormalised.

If a string is to be regarded as made up of discrete partons then it
might make sense to consider the statistics of each parton. In the two
dimensional worldsheet of the string a parton could have fractional
statistics. If string partons are such that an increasing number of
them are seen in a string at higher energies it may be necessary for
the statistics to be divided up into fractions of ordinary fermionic
or bosonic statistics. In the higher dimensional target space only
half integer multiple statistics are permitted to be observed.

Heuristically we might picture the string as an object consisting of
$n$ partons each with an interchange phase factor $q$ such that
$q^n$ is real, i.e. $1$ or $-1$. This suggests that a continuum limit
might exist where $n \rightarrow \infty$ on the worldsheet while the
string has discrete aspects in target space. Such a model might
be based on quantum group symmetries. There are already some encouraging
results which suggest that it might be possible to formulate fractional
superstring models \cite{ArTy91}.

This section would not be complete without referring to a number of other
attempts to understand discrete string theory
\cite{GePa87,FiIs89,FiIs90,AlBa91,FiGaIs92}.


\section*{Discrete Open String Associative Algebras}

To begin constructing event-symmetric string field theories we will extend
matrix algebras to discrete string algebras then use these to construct
extensions of matrix models \cite{Gib94b}.

A basis for a discrete open string vector space is defined by the set of
open ended oriented strings passing through a sequence of events in an
event-symmetric space-time of $N$ events. E.g. a possible basis element might
be written,
\begin{equation}
              C = (1,4,3,1,7)
\end{equation}
Note that a string is allowed to intersect itself. In the example the string
passes through the event $1$ twice. Strings of length one and a zero length
null string can be included in the basis. The order in which the string
passes through the events is significant e.g.
\begin{equation}
(1,4,3) \neq (1,3,4)
\end{equation}
but the order in which the events themselves have been numbered is
irrelevant since the models are to be event-symmetric.
A complete set of field variables in an event-symmetric string model would
be an element of this infinite dimensional vector space which could be written
as a sum over strings $C$
\begin{equation}
                     \Phi = \sum \phi^C C
\end{equation}
This can define either a real or complex vector space. To avoid questions
about convergence in some of the definitions that follow it is easiest to
specify that only a finite number of the components can be non-zero. Other
ways of regularising could be used or the sums could be regarded as just
formal expressions.

The vector components on strings of a given length $r$ can be regarded as
the components of a tensor of rank $r$, e.g.
\begin{equation}
          \phi^C = \phi^{( a,b,c )} = \phi^{abc}
\end{equation}
We will always use lower case indices for events and upper case for strings.
A complete vector on the space can be thought of as a sequence of tensors.
\begin{equation}
           \Phi = ( \phi, \phi^a, \phi^{ab}, \phi^{abc}, \ldots )
\end{equation}

The open string vector space can therefore be naturally related to
a tensor algebra. The tensor product is defined for base strings by
concatenating them together e.g.
\begin{equation}
       ( 1,2,3 ) \otimes
       ( 4,5 ) =
       ( 1,2,3,4,5 )
\end{equation}
This defines associative algebras over the reals or complex numbers
which will be denoted by $Tensor(N, {\Bbb R})$ and $Tensor(N, {\Bbb C})$
The zero length string acts as a unit for this product. In terms of
the tensor components the multiplication is given by,
\begin{equation}
                     \Phi = \Phi_1 \otimes \Phi_2
\end{equation}
\begin{equation}
          ( \phi, \phi^a, \phi^{ab}, \phi^{abc}, \ldots ) =
	  ( \phi_1 \phi_2, \phi_1 \phi_2^a + \phi_1^a \phi_2,
	    \phi_1 \phi_2^{ab} + \phi_1^a \phi_2^b + \phi_1^{ab} \phi_2,
	    \phi_1 \phi_2^{abc} + \ldots, \ldots )
\end{equation}
An inner product can be defined.
\begin{equation}
             \Phi_1 \bullet \Phi_2 = \sum \overline{\phi}_1^C \phi_2^C
\end{equation}
[An overline is used to denote complex conjugation in this paper.]

The inner product will prove useful when we wish to define a positive
definite form for an action since,
\begin{equation}
             \Phi \bullet \Phi = \sum |\phi^C|^2 \geq 0
\end{equation}

The tensor product algebra is actually of limited interest here because
it does not look like the product used to generate the universal string
group. To rectify this we will define a different product $AB$ of two strings
in the space. This is formed by joining them when the end of $A$ matches the
beginning of $B$ reversed with the common end events removed. It is necessary
to add together all the ways in which this can be done e.g.
\begin{equation}
(3,1,7) (7,1,5,1) =
(3,5,1) +
(3,1,1,5,1) +
(3,1,7,7,1,5,1)
\end{equation}
The multiplication defined on the base elements is extended to the
whole vector space by linearity. in terms of tensor components we have,
\begin{equation}
                     \Phi = \Phi_1 \Phi_2
\end{equation}
\begin{equation}
          ( \phi, \phi^a, \phi^{ab}, \ldots ) =
   ( \phi_1 \phi_2 + \phi_1^a \phi_2^a + \phi_1^{ab} \phi_2^{ba} + \ldots,
    \phi_1 \phi_2^a + \phi_1^a \phi_2 +  \phi_1^b \phi_2^{ba} +  \ldots,
	     \ldots )
\end{equation}
This algebra is non-trivially associative with the null string acting as a
unit. The algebra over the reals defined in this way is a discrete open string
algebra denoted by $Open(N, {\Bbb R})$ and its complexification by
$Open(N, {\Bbb C})$.

An undesirable feature of this algebra is that two strings which have
no event in common do not commute. This can be cured by taking the tensor
product of $Open(N, {\Bbb R})$ with the matrix algebra $M(N, {\Bbb R})$.
The matrix algebra is spanned by base elements
\begin{equation}
            ( a ; b )
\end{equation}
with a one in row $a$ and column $b$ of the matrix. The semi-colon is used
here to distinguish these from the open strings of length 2. The matrix
product is simply
\begin{equation}
            ( a ; b )
	    ( c ; d ) =
	    \delta_{bc} ( a ; d )
\end{equation}
The tensor product $Open(N, {\Bbb R}) \otimes M(N, {\Bbb R})$ has a basis
of elements
\begin{equation}
            ( a, b,..., c ) \otimes
	    ( d ; e )
\end{equation}
which we shorten to
\begin{equation}
            ( d; a, b,..., c; e )
\end{equation}
The multiplication defined on these elements is similar to that for
the bare strings except that in the case where the last point of the
first string is not the first
point of the second the product is always zero. This ensures that
string models are local, i.e. strings which do not intersect should
not interact directly. E.g.
\begin{equation}
(1;4,3,1;7) (1;6) = 0
\end{equation}
To ensure associativity this rule is balanced with another rule that
if the whole of one of the strings in a product matches, the last event
is not cancelled e.g.
\begin{equation}
(1;3,1;7) (7;1;3) =
(1;3,1,1;3) +
(1;3;3)
\end{equation}
The matrix algebra is a sub-algebra of this string algebra
so we are justified in regarding the algebra as an extension of
the family of algebras on square matrices.

A notation for the string algebras will be adopted which reflects this
relationship. The string
algebras are extensions of the algebras $M(N,{\Bbb R})$ and $M(N,{\Bbb C})$
and will be written $Open(M(N,{\Bbb R}))$ and $Open(M(N,{\Bbb C}))$.

This leads to some more general definitions. Firstly there is no need for
$N$ in either of the two factors to be the same so define,
\begin{equation}
          Open(N,M(L,{\Bbb C})) = Open(N,{\Bbb C}) \otimes M(L,{\Bbb C})
\end{equation}
The open string extension of a general associative algebra {\cal A} over
complex numbers can be defined as,
\begin{equation}
          Open(N,{\cal A}) = Open(N,{\Bbb C}) \otimes {\cal A}
\end{equation}

For the moment we return to the specific case of the extended matrix algebras
to define the generalised trace and adjoints. The trace is best defined as
the trace from the matrix part. I.e. there is a contribution from the
components of length two strings on the matrix diagonal only.
\begin{equation}
Tr(C) = 1 if C = ( i;i ) and = 0 otherwise
\end{equation}
There is an extended trace defined as the sum over components of any
even length string which is palindromic, i.e. the same when reversed e.g.,
\begin{equation}
OTr (1;4,4;1) =
OTr (3;3) = 1
\end{equation}
\begin{equation}
OTr (2;3) =
OTr (1;2;1) = 0
\end{equation}
Both these traces behave like traces should and in particular,
\begin{equation}
                 Tr(\Phi_1\Phi_2) = Tr(\Phi_2\Phi_1)
\end{equation}
\begin{equation}
                 OTr(\Phi_1\Phi_2) = OTr(\Phi_2\Phi_1)
\end{equation}

The orientation reversal of strings will be used to define transposition
denoted with a $T$. e.g.
\begin{equation}
(1;5,3;4)^T = (4;3,5;1)
\end{equation}
The adjoint of a general element of the space, denoted by a dagger, is defined
by transposing each base
element and in the case of the complex space the complex conjugate of the
components is also taken. I.e.
\begin{equation}
                     \Phi^\dagger = \sum \overline{\phi}^C C^T
\end{equation}
The usual relation between adjoints and multiplication holds
\begin{equation}
                     (AB)^T = B^T A^T
\end{equation}
\begin{equation}
                     (\Phi_1 \Phi_2)^\dagger = \Phi_2^\dagger \Phi_1^\dagger
\end{equation}
Finally the inner product can be written in terms of these operations.
\begin{equation}
                     \Phi_1 \bullet \Phi_2 = Tr(\Phi_1^\dagger \Phi_2)
\end{equation}
The adjoint satisfies the necessary conditions to classify the algebra as a
*-algebra.


\section*{The Open String Lie-Algebras}

{}From the associative algebra $Open(M(N,{\Bbb C}))$ an infinite dimensional
Lie algebra can be defined with the Lie product being given by the
anticommutator,
\begin{equation}
                     [ A , B ] = AB - BA
\end{equation}
This product automatically satisfies the Jacobi identity because of the
associativity of the original algebra product,
\begin{equation}
  [ A , [B , C] ] + [ B , [C , A] ] + [ C , [A , B] ] = 0
\end{equation}
With this product the algebra is an infinite dimensional Lie Algebra
and in principle it defines a group by exponentiation. To avoid complications
in this process only the lie-algebras will be considered.

Using the structure constants for the algebra the Lie product can be written
\begin{equation}
   [ A , B ]   =   \sum f^C_{AB} C
\end{equation}
string indices are formally lowered and raised by reversing
the direction of the string so that
\begin{equation}
                       f_{CAB} = f^{C^T}_{AB}
\end{equation}
The three strings C, A and B are said to form a triplet is $f_{ABC}$ is
plus one, and an anti-triplet if it is minus one. They are a triplet
if and only if they are all different and the end of A matches the beginning
of B, the end of B matches the beginning
of C and the end of C matches the beginning of A without any events being
left out or used twice. Anti-triplets are triplets with two of the strings
interchanged. It follows that the structure constants are fully antisymmetric.
\begin{equation}
         f_{ABC} = f_{BCA} = f_{CAB} = -f_{ACB} = -f_{CBA} = -f_{BAC}
\end{equation}
The following important relation is also valid
\begin{equation}
        A \bullet [ B , C ] = [ A , B ] \bullet C = f_{ABC}
\end{equation}
{}From this description of the Lie-product the relationship with Kaku's
Universal String Group is now clear. The only essential difference is that the
group is now defined on a discrete event-symmetric space-time rather than a
continuous one.

Because the Lie-product was defined as the anticommutator on the string
extended matrix algebra $Open(M(N,{\Bbb C}))$ we know that the Lie-algebra
must be an extension of the general linear Lie-algebra. This is confirmed
by the relation,
\begin{equation}
[ ( a; b ) , ( c; d ) ] =
\delta_{bc} ( a; d ) -
\delta_{ad} ( c; b )
\end{equation}
The algebra is therefore given the name $open(gl(N,{\Bbb C}))$. A number of
other extended Lie-algebras follow immediately by using the anti-commutator
of the appropriate extended associative matrix algebras. e.g.
$open(gl(N,{\Bbb R}))$, $open(gl(N,{\Bbb H}))$, $open(N,gl(L,{\Bbb C}))$
and of course $open(N,{\Bbb C})$.

The trace of the matrix algebras can also be used to define the special
subgroups because,
\begin{equation}
                  Tr [A , B] = 0
\end{equation}
The sub-algebra of traceless elements of $open(gl(N,{\Bbb C}))$ will be
denoted by $open(sl(N,{\Bbb C}))$. The Open trace can also be used to
define subgroups. I.e. the elements of $open(gl(N,{\Bbb C}))$ for which
\begin{equation}
                  OTr(\Phi) = 0
\end{equation}
form the sub-algebra $sopen(gl(N,{\Bbb C}))$. There is also an algebra
$sopen(N,{\Bbb C})$ defined in this way and if both traces are used at once we
have $sopen(sl(N,{\Bbb C}))$.

There is an important alternative definition of the special groups for matrix
algebras. Given a Lie Algebra ${\cal L}_0$ a subalgebra is defined as those
elements which are formed from the Lie-product.
\begin{equation}
                  {\cal L}_1 = [ {\cal L}_0 , {\cal L}_0 ]
\end{equation}
If ${\cal L}_0$ is $gl(N,{\Bbb C})$ then ${\cal L}_1$ is $sl(N,{\Bbb C})$.
For the string extended algebras there are many linear invariant operators $O$
which have the tracelike property.
\begin{equation}
                  O [ \Phi_1 , \Phi_2 ] = 0
\end{equation}
so applying the same technique to $open(gl(N,{\Bbb C}))$ might give a
sub-algebra of $sopen(sl(N,{\Bbb C}))$. However, for an infinite dimensional
Lie-Algebra it is not sure that the sequence will end.

Of more importance to event-symmetric string theories are the open string
extensions to the families of Lie-algebras of compact matrix groups $so(N)$,
$u(N)$ and $sp(N)$. These are easy to define with the adjoint operator which
has the property,
\begin{equation}
     [ \Phi_1 , \Phi_2 ]^\dagger = - [ \Phi_1^\dagger , \Phi_2^\dagger ]
\end{equation}
The algebra $open( u(N) )$ is defined as the sub-algebra of
$open(gl(N,{\Bbb C}))$ containing all elements for which
\begin{equation}
                  \Phi^\dagger = - \Phi
\end{equation}
The algebras $open( so(N) )$ and $open( sp(N) )$ are the similarly defined
sub-algebras of $open(gl(N,{\Bbb R}))$ and $open(gl(N,{\Bbb H}))$.

It must be emphasised that all these algebras can be reduced to the product
of the appropriate matrix lie algebra and a non-local open string algebra.


\section*{Statistical and Quantum Models for Open Strings}

To define a model or theory which incorporates the group structures defined
in the previous sections we need to choose a representation and an invariant
action. The obvious representation to choose is the fundamental representation
which takes elements of the lie-algebra. Often it is useful to regard the
algebra as a family of tensor components.
\begin{equation}
   \Phi = \sum \phi^{ab} ( a; b ) +
   \sum \phi^{abc} ( a; b; c ) + ...\ldots
\end{equation}
The infinitesimal transformations are generated by an element $E$ of
the algebra as follows,
\begin{equation}
          \delta \Phi = [ \Phi , E ]
\end{equation}
There are many alternative representations which could equally well be used
to build models but the fundamental representation has the advantage that
there is exactly one component field variable for each degree of symmetry.

The action should be real and must satisfy the locality
principle. It will take a polynomial form in the components of the
representation and in no term must there appear a product of two components
of strings which do not pass through the same event. This rules out the
non-local groups such as $Open(N,{\Bbb C})$ since they have
very few invariants which are local in this sense. The special groups will
also be ruled out since constraints such as $Tr(\Phi) = 0$ can be considered
non-local.

The trace is a source of invariants since
\begin{equation}
          \delta Tr(\Phi) = Tr[ \Phi , E ] = 0
\end{equation}
Furthermore the associative product can be used since the lie-algebra acts
like a differential operator on the extended matrix algebra according to the
Leibnitz rule,
\begin{equation}
	[\Phi_1 \Phi_2 , E ] =
	 [\Phi_1 , E] \Phi_2 + \Phi_1 [\Phi_2 , E]
\end{equation}
So there is an infinite sequence of invariants given by,
\begin{equation}
                 {\cal I}_n = Tr(\Phi^n),   (n = 1,\ldots)
\end{equation}
Another sequence of invariants can be defined using the extended trace and
there are many other possible invariants but for simplicity only these
will be considered.
Any action which is written as a sum of these invariants is consistent with
the locality condition.
\begin{equation}
                 S = \sum g_n{\cal I}_n
\end{equation}
A statistical model has a partition function defined on a real action which is
positive definite or at least bounded below. For the string extended general
linear groups the trace invariants are not positive definite. This problem is
resolved in the same way as it is for matrix models by using the Lie-algebras
of the compact groups for which
\begin{equation}
                    \Phi^\dagger = - \Phi
\end{equation}
Then the even trace invariants can be written,
\begin{equation}
          {\cal I}_{2n} = tr(\Phi^{2 n}) = (-1)^n \Phi^n \bullet \Phi^n
\end{equation}
In particular (summation convention applies)
\begin{equation}
     {\cal I}_2 = \overline{\phi}^{ab}\phi^{ab} +
     \overline{\phi}^{abc}\phi^{abc} +
                    \overline{\phi}^{abcd}\phi^{abcd} + \ldots
\end{equation}
The simplest non-trivial action for a statistical model is therefore
\begin{equation}
   S  =  m \Phi \bullet \Phi + \beta \Phi^2 \bullet \Phi^2
\end{equation}

[It is important to recognise that the model has  an infinite number of
degrees of freedom even for finite $N$. It would be necessary to demonstrate
that it can give a well defined model despite this.]

There are many other possibilities but this is the most immediately interesting
bosonic open string statistical model. It is also possible to construct
fermionic models using representations such as
\begin{equation}
                 \Psi = \sum \psi^C C
\end{equation}
Where the components $\psi^C$ are anticommuting Grassman variables. an action
for this model can be written,
\begin{equation}
   S  =  im \Psi \bullet \Psi  +
   \beta [ \Psi , \Psi ] \bullet [ \Psi , \Psi ]
\end{equation}

The extended trace can also be used to define positive definite actions because
$OTr(\Phi^2)$ is bounded even though it contains non-square terms,
\begin{equation}
             -OTr(\Phi^2) = \overline{\phi}^{ab}\phi^{ab} +
	     2 \overline{\phi}^{abc}\phi^{abc} +
                    \overline{\phi}^{ab}\phi^{bdda} +
		    \overline{\phi}^{bdda}\phi^{ab} +
		    3 \overline{\phi}^{abcd}\phi^{abcd} + \ldots
\end{equation}
The extended trace has the advantage of suppressing longer strings with
higher coefficients so these models may be better behaved than those based
on the ordinary trace.

For quantum models the conditions can be relaxed a little since the action does
not have to be positive definite to give a well defined partition function.
The general linear groups are still ruled out but extended Poincare groups
might be considered as well as the compact groups and the odd trace invariants
could also be valid terms in the action.

For open string models there appear to be many possible gauge groups, many
possible representations and many possible invariants. There are several ways
to generate many more possibilities than have been described here. For example
models of charged strings can be constructed from algebras such as
$open( N, so(10 N) )$. Some further criterion would be needed to select a
good theory. It is possible to speculate that only a small number of these
models would have the desired symmetry breaking features to identify them
as good theories. This might be considered unsatisfactory since it would be
better to have a kinematic reason for selecting the right model rather than a
dynamic one.

Another feature of the open string models which is unsatisfactory is that the
event-symmetry is not unified with the gauge group. It is true that the
extended matrix models include the symmetric group as a subgroup of the matrix
group. However, true event-symmetry is invariance under permutation of events
and although the models above possess this invariance it is not the same as
the symmetric subgroup of the matrix group which acts only on the ends of
the strings.

To see this notice that the symmetric group is represented by permutation
matrices. I.e square matrices $P$ with a single element equal to one in each
row and column and zero everywhere else. The permutation of events
transforms the tensor components e.g,
\begin{equation}
  \phi^{abc} \rightarrow \phi^{def} P^{da} P^{eb} P^{fc}
\end{equation}
The set of permutation matrices form a subgroup of $O(N)$ which is contained
in the open string group and is generated by length two strings. For an
infinitesimal transform generated with length two strings we get
\begin{equation}
                 E = \sum \epsilon^{ab} ( a;b )
\end{equation}
\begin{equation}
                 \delta\Phi = [ \Phi , E ]
\end{equation}
\begin{equation}
                 \delta\phi^{abc} = \phi^{dbc} \epsilon^{da} +
		 \phi^{abf} \epsilon^{fc}
\end{equation}
But if this is exponentiated it gives
\begin{equation}
          \phi^{abc} \rightarrow \phi^{dbf} P^{da} P^{fc}
\end{equation}
I.e. it acts only on the ends of the string without touching the middle.
Soon we shall see how closed string groups correct this fault.


\section*{Supersymmetric String Groups}

An attractive feature of the discrete string groups on event-symmetric
space-time is that supersymmetric versions can be constructed in a very natural
way. This is in contrast to the situation for continuum string field theories
where supersymmetric generalisations are difficult to construct in
a covariant formalism \cite{Ced94}.

The matrix algebras $M(N, {\Bbb R})$ and $M(N, {\Bbb C})$ can be generalised
to superalgebras $M(L|K, {\Bbb R})$ and $M(L|K, {\Bbb C})$
\cite{Cor89}. From these super algebras a number of families of
super Lie-algebras can be constructed of which the most important include
$gl(L|K, {\Bbb R})$, $gl(L|K, {\Bbb C})$, $u(L|K)$, $osp(L|K)$.

It is possible to apply the string extension methods for ordinary algebras
to these superalgebras to construct $Open(N, M(L|K, {\Bbb C}))$,
$open(N, u(L|K))$ etc. This can be improved by first generalising
$Open(N,{\Bbb C})$ to the super-symmetric algebra $Open(L|K, {\Bbb C})$.
To define this algebra it is sufficient to describe a consistent grading
of the base strings into odd and even strings. To do this the events
themselves are given parity so that event-supersymmetric space-time contains
$L$ even events and $K$ odd events. For notational convenience even events will
be labelled with even integers and odd events with odd integers.

The parity of a string is defined to be the total parity of the events it
passes through. The parity
of a string $C$ written $par(C)$ is zero for even strings and one for odd
strings. This defines a grading of the vector space which is
consistent with the associative multiplication since the parity of the
product of two strings is the sum of their parities modulo two
\begin{equation}
                     par(AB) = par(A) + par(B) - 2 par(A) par(B)
\end{equation}
The components of the vectors must be taken from a Grassman algebra
with even (commuting) variables for components of even strings and
odd (anti-commuting) variables for components of odd strings i.e.
\begin{equation}
                      \Xi = \sum \xi^C C
\end{equation}
\begin{equation}
		      \xi^A \xi^B = (-1)^{par(A)par(B)} \xi^B \xi^A
\end{equation}
The real and complex algebras defined in this way are denoted by
$Open(L|K, {\Bbb R})$ and $Open(L|K, {\Bbb C})$. Note that while
$Open(L|0, {\Bbb R})$ is isomorphic to $Open(L, {\Bbb R})$, the
algebra $Open(0|L, {\Bbb R})$ is a super-algebra in which the parity
of a string is the parity of its length. This is in contrast to the
matrix algebras for which $M(L|0, {\Bbb R})$ is the same as
$M(0|L, {\Bbb R})$.

It is now possible to define local super-matrix algebras
$Open(L|K, M(P|Q, {\Bbb C}))$ using the tensor product prescription.
I.e.
\begin{equation}
            Open(L|K, {\cal A}) =  Open(L|K, {\Bbb C}) \otimes {\cal A}
\end{equation}
In the case $P=L$ and $Q=K$ we write simply $Open(M(L|K, {\Bbb R}))$ for
consistency the indices of the matrix algebra are also taken as odd and
even.

The adjoint operator must fulfill the usual relation
\begin{equation}
                (\Xi_1 \Xi_2)^\dagger = \Xi_2^\dagger \Xi_1^\dagger
\end{equation}
This is achieved by modifying the previous definition to include a factor of
$i$ when taking the adjoint of an odd element. This restricts us to the
complex version of the model.
\begin{equation}
                     \Xi^{\dagger} = \sum i^{par(C)} \overline{\xi}^C C^T
\end{equation}

When generalising the definition of trace and extended trace extra sign
factors are needed corresponding to the parity of half the even string. E.g.
\begin{equation}
Tr (2;2) = 1,  Tr (3;3) = -1
\end{equation}
\begin{equation}
OTr (3;5,5;3) =  1,  OTr (1;4,4;1) = -1
\end{equation}

String extended super Lie-algebras can also be constructed for each
of the supersymmetric families of matrix lie-algebras. From the super-algebra
$Open(M(L|K, {\Bbb C}))$ a Lie-product is defined using the anticommutator,
\begin{equation}
               [ \Xi_1 , \Xi_2 ] = \Xi_1 \Xi_2 - \Xi_2  \Xi_1
\end{equation}
Then the lie product for elements of the representation will be anticommuting.
\begin{equation}
                    [ \Xi_1 , \Xi_2 ] = - [ \Xi_2 , \Xi_1 ]
\end{equation}
But because of the commutation/anti-commutation relations on the components
the Lie product of two odd base elements must be symmetric instead of
anti-symmetric. I.e.
\begin{equation}
                    [ A , B ]_{\pm} = AB - (-1)^{par(A)par(B)} BA
\end{equation}
This defines $open(gl(L|K, {\Bbb C}))$.

A representation of a reduced Lie sub-algebra $open(u(L|K))$ is defined as
those elements which satisfy,
\begin{equation}
                     \Xi^\dagger = -\Xi
\end{equation}

The scalar product is now defined by
\begin{equation}
                     \Xi_1 \bullet \Xi_2 = Tr(\Xi_1^\dagger \Xi_2)
\end{equation}

This product is an invariant for the group $open(u(L|K))$ but is not
positive definite because of the extra minus sign in the trace. Only
a quantum model can be defined.

Actions for a model based on this representation are also the same as
before. In general the action may contain any powers in the algebra squared
with the scalar product.
\begin{equation}
   S  =  g_1 \Xi \bullet \Xi + g_2 \Xi^2 \bullet \Xi^2 +
                                 g_3 \Xi^3 \bullet \Xi^3 + \ldots
\end{equation}
This supersymmetric generalisation is an analogue of the supersymmetric
generalisation of matrix models already described.

It is possible that interesting physics exists in these models in
a large $L,K$ double scaling limit with the constants $g_i$ scaled as
functions of $N$.


\section*{Discrete Closed String Groups}

(Please note errata to this section added at end of paper)

In continuum string theory the closed string field theories are often
considered to be of more physical interest but are also harder to construct.
The same applies to event-symmetric closed string models \cite{Gib94c}.

If we are going to follow the procedure which worked for open strings
we would first construct Closed String algebras in which the base
elements are cyclically symmetric. The extensions use a basis of closed
discrete strings which will be written with double brackets to
distinguish them from the open strings. When they are shifted cyclically
a sign is introduced if they are even length i.e.,
\begin{equation}
(( a, b )) = - (( b, a ))
\end{equation}
\begin{equation}
(( a, b, c )) = (( b, c, a ))
\end{equation}
\begin{equation}
(( a, b, c, d )) = - (( b, c, d, a ))
\end{equation}
Generally we can write using the tensor product symbol to mean
joining strings together that,
\begin{equation}
                 A \otimes B = (-1)^{len(A)len(B)} B \otimes A
\end{equation}
Some strings of even length must be excluded because of this sign rule e.g.
\begin{equation}
((1,1)) = ((1,2,3,1,2,3)) = 0
\end{equation}
This at least defines an infinite dimensional vector space.

The base elements might be multiplied by identifying common sequences
in opposite sense within them. E.g.
\begin{equation}
((1,2,3,4)) ((5,3,2,7)) = - ((1,2,2,7,5,4)) +
((1,7,5,3,3,4)) - ((1,7,5,4))
\end{equation}
The sign for such a multiplication is chosen so that when the matching
segments are moved to the end of the first string and the beginning of the
second it is positive. For locality, when two strings have no points in common
the product is zero. Unfortunately this algebra fails to be associative e.g.
\begin{equation}
((1,2)) \{((2,3)) ((3,4,1))\} =
((1,4,1)) - ((2,2,4)) - ((4))
\end{equation}
\begin{equation}
\{((1,2)) ((2,3))\} ((3,4,1)) =
((1,4,1)) - ((3,3,4)) - ((4))
\end{equation}
Because of this non-associativity we can not be sure that defining a
Lie-product as the anticommutator will satisfy the Jacobi identity.
The associativity problem can be partly removed by allowing non-local
terms but a problem arises in particular with cases where an intermediate
product becomes zero because of the cyclic rules e.g. how can the
following product be made associative if $((1,1))$ and
$((2,2))$ are excluded,
\begin{equation}
((1,2)) ((2,1)) ((1,3))
\end{equation}
We could try a different vector space in which the sign factors were
not included in the cyclic rules but this gives a commutative
associative algebra. Commutative algebras are not very interesting
since the Lie algebra defined in terms of there commutators is going
to be abelian.

Despite this discouraging result we can continue on undaunted and try
to find a Lie-algebra on the vector space without the intermediate step
of an associative algebra. All we need to do is find suitable values for
the structure constants which satisfy the required commutation and Jacobi
relations or their graded equivalents. If $A$, $B$ and $C$ are base strings
in the vector space the structure constants are given by
\begin{equation}
                   [A,B]_{\pm} = \sum_C f^C_{AB} C
\end{equation}

first of all we want to make sure that the structure constants only
allow terms in a product which come from combining the two loops of
string with a contiguous piece cancelled out.

If a string $A$ contains a piece $X$ and a string $B$ contains the
same piece reversed, i.e. $X^T$ then since the piece $X$ can be rotated
to the nominal ends of the string we can write,
\begin{equation}
                 A = a \otimes X
\end{equation}
\begin{equation}
		 B = X^T \otimes b
\end{equation}
When the two strings are joined together and the piece X is cancelled out
the result is,
\begin{equation}
             (AB)_X = ((a \otimes X)(X^T \otimes b))_X
	            = a \otimes b
\end{equation}
What this means is that structure constants must be zero unless they take
the form,
\begin{equation}
                 f^{(a \otimes b)}_{(a \otimes X)}{}_{(X^T \otimes b)}
\end{equation}
At this stage we allow for the possibility that any of the pieces $a$, $b$
or $X$ can be zero length.

One further condition which we should impose is that the Lie-algebra must
be event-symmetric in the sense that the structure constants are invariant
under any permutation (i.e. relabelling) of events.

These conditions combined define an interesting mathematical problem to
which we might seek all possible solutions. The answer would include, for
example, the structure constants of the orthogonal matrix Lie-Algebras on
length two strings with all others zero. We are not going to do a complete
analysis here but we find that there is one elegant solution.

It can be checked that,
\begin{equation}
               (AB)_X = (-1)^{len(A) len(B) + len(X)} (BA)_{X^T}
\end{equation}
Where $len(A)$ is the number of events in a string or piece of string.

The graded commutation relation which must be fulfilled is
\begin{equation}
                [ A , B ]_{\pm} = (-1)^{par(A) par(B)} [ B , A ]_{\pm}
\end{equation}
The most obvious way to satisfy this is to define the parity of a
string to be the parity of its length and take
\begin{equation}
    f^{(a \otimes b)}_{(a \otimes X)}{}_{(X^T \otimes b)}  = \theta(X) =
    (1 - (-1)^{len(X)})/2
\end{equation}
In other words we only include contributions from cancellations of odd
length pieces in the strings. This is already a rather fortunate result
since it automatically gets rid of the possibility of two strings not
commuting when they dont have any events in common so we have some locality.
Furthermore it gives the correct relations for the length two string
sub-algebra to be the orthogonal Lie-algebra.

It is not immediately obvious that this is going to also
fulfill the graded Jacobi identities which are,
\begin{equation}
                [ A , B ]_{\pm} = (-1)^{par(A) par(B)} [ B , A ]_{\pm}
\end{equation}
\begin{equation}
     (-1)^{par(A) par(C)} [ [ A , B ]_{\pm} , C ]_{\pm} +
     (-1)^{par(B) par(A)} [ [ B , C ]_{\pm} , A ]_{\pm} +
     (-1)^{par(C) par(B)} [ [ C , A ]_{\pm} , B ]_{\pm} = 0
\end{equation}
but in fact it does. The proof requires a number of different cases to be
considered.

In general the three strings will
have various pieces in common and the expression will expand into a sum
over the cancellation of various combinations. We can simplify matters
by restricting to just various explicit pieces. For example the strings
might decompose as,
\begin{equation}
                    A = a \otimes X^T \otimes b \otimes Y
\end{equation}
\begin{equation}
		    B = c \otimes Y^T
\end{equation}
\begin{equation}
		    C = d \otimes X
\end{equation}
if we only look at contributions from the cancellation of $X$ and $Y$ we
can write,
\begin{equation}
 [ [ A , B ]_{\pm} , C ]_{\pm} = \theta(Y) \theta(X) ((AB)_Y C)_{X^T}
\end{equation}
\begin{equation}
 [ [ B , C ]_{\pm} , A ]_{\pm} = 0
\end{equation}
\begin{equation}
 [ [ C , A ]_{\pm} , B ]_{\pm} = \theta(Y) \theta(X) ((CA)_X B)_Y
\end{equation}

We can show that,
\begin{equation}
         ((CA)_X B)_Y = (C (AB)_Y)_X
	         = (-1)^{len(C)(len(A) + len(B)) + len(X)} ((AB)_Y C)_{X^T}
\end{equation}
It can now be seen that the Jacobi identity is satisfied for these
contributions given our choice of structure constants.

This covers many, but not all, contributions to the Jacobi identity.
The above cancellations would also have been found for a wide range
of other possible structure constants including those of the form
\begin{equation}
    f^{(a \otimes b)}_{(a \otimes X)}{}_{(X^T \otimes b)}
       = \theta(X) \sigma(len(X))
\end{equation}
for any arbitrary function $\sigma(l)$. There are other contributions from
cases where the strings decompose as follows,
\begin{equation}
                    A = a \otimes X^T \otimes Y
\end{equation}
\begin{equation}
		    B = b \otimes Y^T \otimes Z
\end{equation}
\begin{equation}
		    C = c \otimes Z^T \otimes X
\end{equation}
Additional terms appear when $X$ is even length and $Y$ and $Z$
are odd length are,
\begin{equation}
 [ [ A , B ]_{\pm} , C ]_{\pm} +=
        ((AB)_Y C)_{X^T \otimes Z} \sigma(len(Y)) \sigma(len(X) + len(Z))
\end{equation}
\begin{equation}
 [ [ B , C ]_{\pm} , A ]_{\pm} +=
        ((BC)_Z A)_{Y^T \otimes X} \sigma(len(Z)) \sigma(len(Y) + len(X))
\end{equation}
These will cancel in the Jacobi formula provided
\begin{equation}
       \sigma(2l-1) \sigma(2k+2m-1) = \sigma(2m-1) \sigma(k+2l-1)
\end{equation}
for all positive integers $k$, $l$ and $m$. The most general solution is,
\begin{equation}
       \sigma(2l-1) = r p^l
\end{equation}
where $r$ and $p$ are arbitrary numbers. If $r = 0$ the algebra is abelian.
if $r \neq 0$ then the base elements can be rescaled by a factor r so it
is isomorphic to the $r = 1$ case. If now $p = 0$ then we have just the
orthogonal lie algebra on length two strings. If $p \neq 0$ then we can
rescale all base elements by a factor $p^{l/2}$ so it is isomorphic to
the $p = 1$. This last step only works for the complex algebra. In the
real case we are left with two possibilities with $p = \pm1$.

There is one possible flaw in this construction, namely in the case where
the product of two of the strings have terms which must be set to zero
because of the cyclic rules. This spoilt our attempts to construct an
associative algebra but by checking cases it can be verified that the
Lie-algebras are not spoilt by this. Note however that we must include
length one strings in the algebra and that, unlike the open string case,
we must include terms where the whole of an odd length string cancels with
part of the other in a product. We can choose to include a zero length
string if we want to and we shall. It commutes with all others but appears
in the product of an odd length string with its transpose.

We can now conclude that we have correctly identified some rather non-trivial
infinite dimensional Lie-algebras defined on the vector space of discrete
loops in event-symmetric space-time. The real and complex Lie-algebras are
given the names $closed_{\pm}(0|N,{\Bbb R})$ and $closed(0|N,{\Bbb C})$
respectively. The sign subscript on the real algebra is the sign of the
parameter $p$ above and will usually be left out and assumed to be $+$.
The sub-algebras generated by the
length two strings are $so(0|N,{\Bbb R})$ and $so(N,{\Bbb C})$.

An encouraging feature of these closed string Lie-algebras is that the
event-symmetry
is included in the algebra in a sense that was not true for the open
string algebras. This is because the matrix sub-algebra acts on all parts
of the string in the appropriate fashion for the representation to be
considered as a family of tensor representations of the matrix algebra.
\begin{equation}
       \Xi = \xi(( )) +
       \sum \xi^{a}(( a )) +
       \sum \xi^{ab}(( a,b )) +
       \sum \xi^{abc}(( a,b,c )) + \ldots
\end{equation}
\begin{equation}
       \xi^{ab} = - \xi^{ba}
\end{equation}
\begin{equation}
       \xi^{abc} = \xi^{bca} = \xi^{cab}
\end{equation}
\begin{equation}
	     etc.
\end{equation}
The components of odd length strings are, of course, anticommuting Grassman
variables. A small change generated by the matrix sub-algebra gives e.g.
\begin{equation}
                E = \sum \epsilon^{ab} (( a,b ))
\end{equation}
\begin{equation}
  \delta \xi^{abc} = \sum_d (\xi^{dbc} \epsilon^{da} +
  \xi^{adc} \epsilon^{db} + \xi^{abd} \epsilon^{dc} )
\end{equation}
This is the correct transformation law for $\xi^{dbc}$ as a third rank
tensor under the group $SO(N)$ generated by $\epsilon^{dc}$. The corresponding
transformation for the open string lacks the middle term. The higher rank
components also transform correctly for the closed strings.
The Alternating group $A(N)$ is a sub-group of $SO(N)$ and acts to permute
events. Because of this the closed string algebra can be said to unify
space-time symmetries and gauge symmetries in a unique and powerful way.

It is possible to go further and say that the discrete string symmetries
give an idea of what is meant by stringy space-time or the geometry of
strings \cite{FrGa93}. A space-time event can be considered as a discrete
string of length one. Strings of length two generate transformations of
space-time analogous to diffeomorphisms on continuous space-time. In general
longer strings generate transformations which intermix strings of different
lengths.

In the event-symmetric open string models this unification appears to be
absent. This can be corrected by defining string groups which include
the closed strings and open strings together. This observation is perhaps
related to the fact that continuum open string theories must necessarily
include closed strings.

The adjoint operator can be defined in the usual way for supersymmetric
adjoints on the complex algebra.
\begin{equation}
                     \Xi^{\dagger} = \sum i^{par(C)} \overline{\xi}^C C^T
\end{equation}
The transpose of a string is its reversal. There is no ambiguity about which
event it is transposed because of the cyclic relations.
The sub-algebra of elements which are equal
to minus their adjoints can be taken and will be denoted by simply
$closed(0|N)$. This is an algebra of non-oriented closed discrete
super-strings.

To complete the construction of an event-symmetric closed string field theory
some invariants must be found. Because the components of the fundamental
representation transform as a family of tensors under the orthogonal matrix
subgroup it is necessary that any invariant must be written as contractions
over indices of tensor products. This condition is not sufficient however.

First of all we should look for a quadratic invariant and can try,
\begin{equation}
                I(\Xi) = \sum q(len(C)) \xi^{C^T} \xi^C
\end{equation}
with the a form factor $q(r)$ depending only on the length of the strings
(i.e. the rank of the tensor) to be determined. However, the odd terms are
identically zero due to anti-commutivity and the even part with $q(l) = 1$
is only invariant for the bosonic sub-group generated by even length strings.

The problem of finding invariants can be solved by using the adjoint matrix
representation. For each element $\Xi$ an infinite super-matrix $M(\Xi)$
acting on the graded vector space of the algebra is defined with components,
\begin{equation}
                  M(\Xi)^A_B = \sum \xi^C f^A_{BC}
\end{equation}
If we consider only elements $\Xi$ with only a finite number of non-zero
components then the matrix has an infinite number of non-zero components
but only a finite number in each row or column. Therefore the product of
any number of these matrices is well defined.

These matrices form a representation of the super Lie-algebra with the
graded anti-commutator as the Lie-product. Invariants can therefore be
constructed using trace and product.
\begin{equation}
                  Tr(M(\Xi)) = \sum M(\Xi)^C_C
\end{equation}
\begin{equation}
                  I_n(\Xi) = Tr(M(\Xi)^n)
\end{equation}
The first invariant $I_1(\Xi)$ receives contributions from components of even
length palindromic strings but there is a regularisation problem. The sum
includes each contribution an infinite number of times. We can however
renormalise this to a definition of an extended trace e.g.,
\begin{equation}
                  CTr ((1,2,2,1)) = 1
\end{equation}
\begin{equation}
		  CTr ((1,2,3,3,2,1)) = -1
\end{equation}
Similar procedure can be applied to the higher order invariants and
with these it is possible to define event-symmetric quantum closed
superstring field theories.

What about the algebras $closed(K|L,{\Bbb C})$, etc., how should they be
defined? It is possible to generalise to these algebras by grading events
as was done for the open strings. It is necessary to redefine the cyclic
conditions on the base elements so that only the permutation of odd events
effect the sign. The parity of a string is then the sum of the parities of
the events it passes through. The algebras defined in this way are less
interesting to us since they are not really event-symmetric. If all events
are taken as even then the algebra is commutative.


\section*{Simplex Groups}

Another class of groups closely related to the string groups is
based on sets of discrete events where the order does not matter
accept for a sign factor which changes according to the signature
of permutations,
\begin{equation}
(( a | b | c )) = -(( b | a | c ))
\end{equation}
\begin{equation}
	etc.
\end{equation}
A base element of length $n$ can be associated with a $n$-simplex
with vertices on the events in the element.

Single event simplices $(( a ))$ and a null simplex
$(( ))$ are included in the algebra.

Multiply by cancelling out any common events with appropriate sign factors.
To get the sign right, permute the events until the common ones are at the
end of the first set and at the start of the second in the opposite
sense. The elements can now be multiplied with the same rule as for the
open string. The same parity rules as for closed string apply. I.e. only
cancellations of an odd number of events is permitted.

The Lie product of two base elements can only be non-zero if they have an odd
number of events in common. e.g.
\begin{equation}
[ (( 1 | 2 | 3 )) (( 2 | 3 | 4 )) ]_- = 0
\end{equation}
\begin{equation}
[ (( 1 | 2 | 3 | 4 )) (( 4 | 3 | 2 | 5 )) ]_+ = (( 1 | 5 ))
\end{equation}
This defines real and complex super-lie algebras which will be called
$simplex(0|N, {\Bbb R})$ and $simplex(0|N, {\Bbb C})$. These Lie algebras
are finite dimensional with dimension $2^N$.

An adjoint can be defined on the complex super-algebra in the usual way
\begin{equation}
                  \Xi = \sum \xi^C C
\end{equation}
\begin{equation}
           \Xi^{\dagger} = \sum \overline{\xi}^C i^{par(C)} C^T
\end{equation}
\begin{equation}
                              = \sum \overline{\xi}^C i^{len(C)} C
\end{equation}

If we take the sub-algebra of elements of $simplex(0|N, {\Bbb C})$ for which
\begin{equation}
                   \Xi^{\dagger} = - \Xi
\end{equation}
then this can be written in terms of their components as
\begin{equation}
                     \overline{\xi}^C = -i^{len(C)} \xi^C
\end{equation}
So
\begin{equation}
                    \xi^C = \phi^C i exp(i [\pi / 4] len(C))
\end{equation}
        With $\phi^C$ being real. If we use these as components writing,
\begin{equation}
                        \Xi = \sum \phi^C C_R
\end{equation}
\begin{equation}
                    C_R = i exp(- i [\pi / 4] len(C)) C
\end{equation}
It can be checked that the basis on $C_R$ has the same multiplication rules
as the basis on $C$ except for an extra minus sign when the number of
common events cancelled is 3 mod 4 just as in the algebra
$closed_{\pm}(0|N,{\Bbb R})$. This is the group $simplex(0|N)$.

The representations of these groups are families of fully antisymmetric
tensors. The Lie algebras are finite dimensional and it is therefore an
interesting exercise to determine how they correspond to the classification
of semi-simple Lie-algebras by factorising into well known compact groups.

An important remark about the simplex groups is that they have a resemblance
to the event-symmetric spinor models which can be seen when their components
are written as families of alternating tensors. In fact it is not
difficult to see that they are generated by the Clifford algebras for
$N$ dimensional space.

A matrix representation of the algebra can be constructed using Gamma matrices
which have size $2^{N/2} \times 2^{N/2}$ provided $n$ is even. In this
representation a mapping between the basic elements is defined by
\begin{equation}
(( a )) \rightarrow \gamma_a
\end{equation}
The gamma matrices satisfy the anticommutation relations,
\begin{equation}
            \gamma_a \gamma_b + \gamma_b \gamma_a = 2 \delta_{ab}
\end{equation}
The full algebra is generated from the linear span of all $2^N$ possible
products of the matrices e.g.
\begin{equation}
(( a | b )) \rightarrow \gamma_a \gamma_b
\end{equation}
The null simplex maps onto the identity matrix.

Since these are all linearly independent matrices with $2^N$ matrix elements
it follows that the algebra over the complex numbers is isomorphic to the
full matrix algebra $M(2^{N/2}, {\Bbb C})$. However, we are interested in
the $Z_2$ graded algebra where the parity is given by the size of the
simplex. It is possible to construct the gamma matrices so that they all have
elements in only the upper right and bottom left quadrants. The grading then
maps the algebra onto the super matrix algebra $M(L|L,{\Bbb C})$, where
$L = 2^{N/2-1}$. It follows that the Lie-superalgebra formed
{}from the graded anticommutators is just the super-symmetric affine
algebra and,
\begin{equation}
            simplex( N, {\Bbb C} ) \simeq gl( L|L, {\Bbb C} )
\end{equation}
while the adjoint defined on the signature algebra corresponds to the usual
adjoint on supermatrices so,
\begin{equation}
            simplex( N ) \simeq u( L|L )
\end{equation}
{}From this it is possible to construct and understand the invariants of the
algebra as invariants of the matrix super-groups. These are functions of the
supertrace of powers of the matrices.

The first order invariant turns out not to be the component corresponding
to the null simplex as you would expect. Instead it corresponds to the
simplex formed from all the $N$ events,
\begin{equation}
             U = (( 1,2,...,N ))
\end{equation}
This and higher order invariants seem to have anything but a local
nature since they are sums over products of simplices which include all
events but which have no event in common.

It is interesting to compare this incomplete study of symmetries on
simplices with earlier work of a similar nature.
Finkelstein also noted the importance of Clifford algebras in
this context \cite{Fin82,FiRo84,FiRo85}. The ideas presented here were derived
independently but the concurrence is important.
It is possible that the supersymmetry described here might lead to
further developments in this area.


\section*{Multi-loop String Algebra}

The closed string group and the simplex group might be both sub-groups
of a larger multi-loop group. The base elements of this group would represent
sets of closed loops. The closed loop group would be the subgroup of single
loops in the multi-loop group and the simplex groups could correspond to sets
of loops each containing exactly one point.

The notation is chosen to be consistent with the loop and simplex groups.
For example a double loop base element with one loop of length three and one
of two would be,
\begin{equation}
(( a,b,c|d,e)) = - (( a,b,c|e,d )) =
(( b,c,a|d,e)) = (( d,e|a,b,c )) etc.
\end{equation}
The sign factor is always the signature of the permutation on the events
in the string.
Antisymmetric multiplication is the obvious generalisation of multiplication
on the closed and simplex groups.

However, this proposed group has some difficulties in construction and it
is not clear that a consistent local multiplication can be defined. A
more satisfactory interpretation of multi-loop states may arise in the
universal enveloping algebra of the closed loop Lie-Algebra. This algebra
has elements which are products of any number of closed loops modulo
the commutation relations of the algebra.


\section*{Deformation?}

We have seen various ways in which we can construct symmetries for discrete
string theories. The closed loop algebras are especially elegant. The
difficulty is that models constructed to have these symmetries are not
well defined.

If a perturbation theory is calculated for the closed loop
theory with a 4-string interaction the loops form faces of tetrahedrons which
in the perturbation theory would seem to construct simplicial complexes in
a fashion similar to tensor models of 3-d gravity. The problem is that there
is no control over the length of the strings. Summations are unrestricted
and divergent. The situation looks like the phase space of the
Ponzano-Regge model \cite{PoRe68} and it is natural to wonder if a quantum
group deformation could be found to regularise the discrete string models in
the same way as the Turaev-Viro model \cite{TuVi92} does for Ponzano-Regge.

In quantum group representation theory the case where $q$ is an $n$th root
of unity the number of principle representations is limited to a finite
size. In the $SU_q(2)$ case the $j$ numbers are limited to $j<n$ unlike
the classical group $SU(2)$ which has representations for any integer $j$.
The expected consequence for a deformed discrete string group would be
that the length of strings has a finite limit controlled by $n$ and that
finite models could thus be constructed.

There are a number of possible avenues which could lead to a deformed
version of the discrete string groups. One approach would be to start
{}from the Universal enveloping algebra of the closed string Lie Algebra.
The basis for this algebra can be regarded as sets of loops. It is
possible that in the deformed model these would have to be replaced by
links whose knottings are significant. If we start from the simpler
simplex algebras then the permutations of events would naturally be
replaced by braidings and the sign factors would be replaced by $q$
values. Multiplication would then be seen as tying together the
ends of strings and it seems possible that this would force a
picture in which an algebra of tangles played an important role
\cite{Bae92}. Unfortunately this approach seems to lead to tangles in
more ways than one.

In a different direction we could look towards loop groups defined as
a form of affine Kac-Moody algebra \cite{PrSe88}. These loop groups are
functions from a circle to a group made invariant under
reparameterisation. Discrete versions would be discrete closed loops
indexed by a Lie group. The Lie group can be deformed to a quantum
group such as $SL_q(N)$. Models on such loops could be made analogous to
the discrete groups described here which would make them similar to
the kind of model used by Boulatov as a pregeometric construction of
lattice topological field theories \cite{Bou92a,Bou92b,Bou93}.

To see this in more detail recall that Boulatov uses functions of 3
group variables with a cyclic condition,
\begin{equation}
                 \phi(g_1,g_2,g_3) = \phi(g_2,g_3,g_1)
\end{equation}
and
\begin{equation}
                 \phi(g_1,g_2,g_3) = \overline{\phi}(g_2,g_1,g_3)
\end{equation}
and
\begin{equation}
                 \phi(g_1,g_2,g_3) = \phi(g_2 g, g_1 g, g_3 g)
\end{equation}
these can be compared with a loop group defined as a discrete loop
of three points mapped onto the group and taken modulo the group $G$.
An action is then formed by taking a tetrahedron vertex and the group
$G$ can be conveniently generalised to a quantum group using the
appropriate intertwiners to generalise the above relations. The
perturbation theory for the action formally generates 3 dimensional
Lattice Topological Field Theories.

Generalisations to functions of four group variables can generate 4D
models but the result seems not sufficiently general to be interesting.
Suppose now we define an algebra analogous to our supersymmetric
simplex algebras using the same formalism. This would have a family
of functions of any number of group variables. A simplex group model
then translates into an action of the type given by Boulatov but
with a new symmetry structure. The perturbation theory will be complex
but would seem to combine simplices of all dimensions into some
generalisation of a TLFT. Again the simplex group could be replaced
by loop groups giving a different result.

In such models there would be two regularisation variables: the number
of events $N$ which is the size of the underlying quantum group and
the deformation parameter $q$. It is plausible that a non-trivial limit
exists by letting $q \rightarrow 1$ at $n$th roots of unity. In this
case the statistics of each event in a string are braided but $n/2$ of them
together can have fermionic statistics. The string could be interpreted
as a string made up of "wee partons" in the sense proposed by Susskind
\cite{SuGr94}. A spacing $a$ between string events could be scaled as
$q \rightarrow 1$ in such a way that $an$ remains a finite length which
is the minimum observable length of the theory. If such a model could be
realised it might explain the paradoxical continuum-discrete dual nature
of strings.

The simplex groups may be significant in understanding how such a model
can lead to a space-time construction in some low-energy limit. The simplex
groups could be seen as reductions of the string groups. Deformation
of the simplex groups may be realised through deformation of Clifford
algebras \cite{BrPaRe93}.

In $q$-deformed models we retain the concept of event-symmetric space-time
but the symmetric group $S(N)$ is replaced with the braid group $B(N)$.

Such ideas are very speculative and it is not clear how such a model, if
it exists, could make contact with string theory as we currently understand
it. One possibility is that the algebraic multiplication rules could be
related to the fusion rules of a conformal field theory \cite{Fuc92,Fuc93}.
Duality transformations must also be important since it is believed that
string theories possess a complex duality structure which combines
target space dualities and coupling dualities \cite{HuTo94}.


\section*{Discussion}

My principle argument in this paper is that we must appeal to symmetry
as our main guide. We know that diffeomorphism invariance and internal
gauge symmetry are among the most fundamental principles in the
standard models of physics. String theory suggests to us that these
should be unified in some much larger symmetry which is broken at
low energies. Conventional wisdom among string theorists may be that
this can be understood in terms of topological quantum field theory.
The successful quantisation of three dimensional gravity shows that
many of the different approaches to quantum gravity, including string
theory, loop representations and simplicial lattice models can indeed be
combined into a topological quantum field theory. This must be
significant.

It seems most credible that a unified theory will incorporate a sum
over different space-time topologies and will allow for topology
change in physics. The symmetry group for diffeomorphism invariance
on different topologies are not the same but if we are to formulate
a theory which has a symmetry as a fundamental rule then that symmetry
must either contain the symmetry of each manifold as a sub-symmetry
or provide a homomorphism onto them. It seems to be an almost inescapable
conclusion that the only way to achieve this is to extend the symmetry
to include the group of all one-to-one mappings on a manifold without
imposing any differentiability or continuity condition. This gives us
the symmetric group of the manifold and resolves the dilemma because
the symmetric group on any two sets are isomorphic if and only if the
two sets have the same cardinality. This I have called event-symmetric
space-time and I have gone on to investigate ways in which the
symmetric group can be further enlarged to include various gauge
symmetries and, in particular, the gauge symmetry of string theories.
I believe that it is a symmetry of this type which describes the full
unbroken symmetry of string theory which is restored beyond the Planck
energy scale. This symmetry must therefore be a symmetry of the
topological phase.

The theory is incomplete and may yet require additional elements. The use
of quantum groups in a deformed version of the theory would bring in
principles from non-commutative geometry as well as knot
theory which is believed to be an important feature of quantum gravity.
A q-deformed theory may lead to finite models with a possible resolution
of discrete/continuum duality.


\section*{Errata}

There is an important error in my proof of the Jacobi Identity for the
closed string Lie-superalgebra and in fact the Identity does not hold
in the following case:
\begin{equation}
A = ((1,2,3,4,5))
\end{equation}
\begin{equation}
B = ((1,4,3))
\end{equation}
\begin{equation}
C = ((5))
\end{equation}
I am grateful to R. Borcherds for kindly providing this counter example.

It seems that the only useful way to correct this anomaly is to redefine
the Lie-superalgebra in such a way that commutators of single loops can give
terms including multi-loops. My effort to construct this correction has
already led to interesting new ideas which I intend to report on in the
near future.

Please send your comments and corrections
by e-mail to phil@galilee.eurocontrol.fr


{\LARGE {\bf Acknowledgements}}

I would like to express my gratitude to those who have made the physics
literature accessible through multimedia on the internet. Special thanks
must go to Paul Ginsparg for setting up the physics e-print archives at
Los Alomos. Without the success of such a service my research would not
have been possible. I am also deeply indebted to the librarians at
SLAC, DESY and CERN for making the databases such as QSPIRES available
on the World Wide Web. My work has greatly benefited from the facilities
of the references and postscript databases.

A special thanks to the Biblioteque Interuniversaire Physique Recherche
at Jessieu for permitting me free access to journals in more conventional
form. My gratitude is also extended to Eurocontrol for providing me with
access to the internet and allowing me to access it during my lunch breaks
{}from work.

Finally, I would like to thank Leonard Susskind for his encouraging e-mail
messages.



\end{document}